\documentclass[epjST]{svjour}

\pdfoutput=1
\usepackage{amsmath}
\usepackage{amssymb}
\usepackage{slashed}
\usepackage{graphicx}
\usepackage{graphics}
\usepackage{epsfig}
\usepackage{color}
\usepackage{xcolor}
\usepackage{cancel}
\usepackage{soul}
\usepackage{hyperref}
\usepackage[section]{placeins}
\usepackage{mathrsfs}
\usepackage{dsfont}
\usepackage[normalem]{ulem}
\usepackage{comment}
\usepackage{mathtools}

\usepackage{subcaption}

\newcommand{\be}{\begin{equation}}
\newcommand{\ee}{\end{equation}}
\newcommand{\bea}{\begin{eqnarray}}
\newcommand{\eea}{\end{eqnarray}}

\def \nn{\nonumber}

\begin{document}

\title{Higher-order curvature operators in causal set quantum gravity
}

\author{
	Gustavo. P.~de Brito\inst{1} \fnmsep\thanks{\email{gustavo@cp3.sdu.dk}} \and 
	Astrid Eichhorn\inst{1}\fnmsep\thanks{\email{eichhorn@cp3.sdu.dk}} \and 
	Christopher Pfeiffer\inst{1,2}
}
\institute{
CP3-Origins, University of Southern Denmark, Campusvej 55, DK-5230 Odense M, Denmark \and 
Niels Bohr Institute, University of Copenhagen, Blegdamsvej 17, 2100 Copenhagen, Denmark
}

\abstract{
We construct higher-order curvature invariants in causal set quantum gravity. The motivation for this work is twofold: first, to characterize causal sets, discrete operators that encode geometric information on the emergent spacetime manifold, e.g., its curvature invariants, are indispensable. Second, to make contact with the asymptotic-safety approach to quantum gravity in Lorentzian signature and find a second-order phase transition in the phase diagram for causal sets, going beyond the discrete analogue of the Einstein-Hilbert action may be critical.\\
Therefore, we generalize the discrete d'Alembertian, which encodes the Ricci scalar, to higher orders. We prove that curvature invariants of the form $R^2 -2 \Box R$ (and similar invariants at higher powers of derivatives) arise in the continuum limit.
}

\maketitle

\section{Introduction and motivation}
In this paper, we construct higher-order curvature operators for causal sets. Our motivation is twofold: 
First, geometric quantities such as curvature operators are important when one reconstructs a continuous spacetime from a discrete causal set -- which is one of the key outstanding problems in causal set quantum gravity. Second, higher-order curvature operators are important when one uses causal sets to search for asymptotic safety in quantum gravity in Lorentzian signature -- which is one of the key outstanding problems in asymptotically safe quantum gravity.\\
Below, we introduce these motivations in more detail.\\

Our first motivation comes from the reconstruction of continuum geometry from a discrete causal set.
Causal set quantum gravity is based on a discretization of Lorentzian spacetimes \cite{Bombelli:1987aa}, see \cite{Surya:2019ndm} for a recent review.  It substitutes Lorentzian continuum manifolds by networks of spacetime points, in which the links that connect the nodes of the network correspond to causal relations. Mathematically, such networks are partial orders. However, the set of partial orders which satisfy the causal-set requirements is much larger than the set of partial orders that encode Lorentzian manifolds \cite{Henson:2015fha}.\\
 Thus, a central challenge in causal set quantum gravity is  to reconstruct continuum information from a causal set. We thus require geometric quantities  to characterize a causal set and decide, whether or not a given causal set  is the discrete counterpart of a continuum manifold, and if so, of which manifold.  These quantities must be calculable purely in terms of the causal relations making up the causal set.
Already existing geometric and topological quantities include measures of the spacetime dimensionality \cite{Myrheim:1978ce,Reid:2002sj,Eichhorn:2013ova,Glaser:2013pca,Roy:2013} as well as spatial dimensionality on antichains (the causal-set analogue of a spatial hypersurface) \cite{Eichhorn:2019uct}, measures of geodesic distance \cite{Brightwell:1990ha,Rideout:2008rk,Eichhorn:2018doy},  and a construction of the Ricci curvature scalar \cite{Benincasa:2010ac}, for further examples, see the review \cite{Surya:2019ndm}. Here, we will generalize the curvature scalar to higher orders and to derivatives of the curvature. This provides further diagnostics to characterize causal sets.\\

Our second motivation comes from the asymptotic-safety paradigm in Lorentzian-signature spacetimes.  In the asymptotic-safety paradigm, gravity is quantized as a predictive quantum field theory \cite{Reuter:1996cp,Souma:1999at,Lauscher:2001ya,Reuter:2001ag}. This is possible, if this quantum field theory starts out in a quantum scale-invariant regime at very small distance scales, see, e.g., the recent reviews \cite{Eichhorn:2018yfc,Pereira:2019dbn,Eichhorn:2020mte,Pawlowski:2020qer,Reichert:2020mja,Bonanno:2020bil,Eichhorn:2022jqj,Eichhorn:2022gku}.\\
A main challenge is that, with few exception \cite{Manrique:2011jc,Fehre:2021eob}, calculations are done in  Euclidean signature. The functional Renormalization Group method \cite{Wetterich:1992yh}, see \cite{Dupuis:2020fhh} for a review, which these calculations use, can be extended to Lorentzian signature, but at the cost of significant technical challenges. Thus, a different framework to investigate the asymptotic-safety paradigm in Lorentzian signature is called for. In \cite{Eichhorn:2017bwe,Eichhorn:2019xav}, it was proposed that causal sets could provide such a framework. Following the proposal, the discreteness in causal sets is viewed as a regularization of the quantum gravitational path integral. A universal continuum limit can be taken, sending the regularization scale to zero, if a second-order phase transition exists in the phase diagram of causal sets.  This same idea is explored in causal dynamical triangulations \cite{Ambjorn:2012ij,Ambjorn:2014gsa,Ambjorn:2020rcn} and in other combinatorial approaches \cite{Trugenberger:2016viw,Bahr:2016hwc}.
First investigations of the phase diagram for causal sets exist, but show only first-order transitions if the action that is used contains only a curvature term \cite{Surya:2011du,Glaser:2017sbe,Cunningham:2019rob}. There are also indications that the presence of matter could induce new phase transitions, albeit not higher-order ones \cite{Glaser:2020yfy}. Interestingly, Euclidean continuum studies suggest that an asymptotically safe regime has more than two relevant directions, which need to be tuned in order to reach the corresponding phase transition \cite{Lauscher:2002sq,Machado:2007ea,Codello:2008vh,Benedetti:2009rx,Benedetti:2012dx,Falls:2014tra,Falls:2018ylp,Falls:2020qhj}. This suggests that it may be necessary to go beyond an action containing only a single curvature term in order to find a higher-order phase transition. This motivates us to construct higher-order curvature terms for causal sets.\\

This paper is organized as follows: In Sec.~\ref{sec:causalsets} we introduce key features of causal sets, before constructing a higher-curvature operator in Sec.~\ref{sec:constructionR2}, where we also investigate its continuum limit.  We then generalize the operator to higher orders in the curvature in Sec.~\ref{sec:higher_order_operators} and discuss the causal-set action with higher-order terms in Sec.~\ref{sec:action}, before we conclude in Sec.~\ref{sec:conclusions}. Additional technical details are provided in an appendix.

\section{Lightning review of causal sets}\label{sec:causalsets}
The causal-set approach to quantum gravity is based on the Hawking-Malament theorem \cite{Hawking:1976fe,Malament:1977}, which states that, under some quite generic assumptions, the causal order contains all information in the metric except for the conformal factor. The causal order of spacetime points $x,y$ is a partial order: For spacetime points at timelike or null separation, $x \preceq y$ (x precedes y), if $x$ is in the causal past of $y$.\footnote{Note that $x \preceq x$.}
For spacetime points at spacelike separation, there is no ordering. \\
A causal set $C$ is a set of spacetime points, together with the relation $\preceq$, such that the following holds:
\begin{enumerate}
	\item[(i)] If  $x \preceq y$ and $y \preceq z$ (and $x \neq z$), then $x \prec z$ (transitivity).
	\item[(ii)] If $x \preceq y$ and $y \preceq x$, then $x = y$ (acyclicity or no closed-timelike-curves)
	\item[(iii)] $|\{z \in C: x\preceq z \preceq y\}|< \infty$ (local finiteness).
\end{enumerate} 
The first two conditions hold for the causal order of spacetime points in any causal
 manifold. The last condition imposes spacetime discreteness, because it restricts the causal interval between any two points to finite cardinality.\\
 Because of the discreteness, one can recover the additional piece of information to reconstruct the metric, namely the conformal factor: by counting the elements in any subset of the causal set, one obtains the volume, because on average each spacetime point $x \in C$ is associated with a spacetime volume $\ell^d$, where $d$ is the spacetime dimensionality and $\ell$ is the discreteness scale. The association is a statistical one: spacetime points are associated with spacetime volumes through a probability distribution, namely a Poisson distribution, which says that the probability to find $n$ spacetime points in a volume $V$ is given by
\begin{eqnarray}
	p(n, V) = \frac{1}{n!} \left(\rho V \right)^n\, e^{-\rho V},\label{eq:Poisson}
\end{eqnarray}
where $\rho = \ell^{-d}$ is the density. This statistical association circumvents a problem that regular, lattice-like discretizations have, namely the breaking of local Lorentz invariance \cite{Bombelli:2006nm}.
\\
Causal sets are the basis for a non-quantum-field theoretic quantization of gravity, in which the metric is discarded at the fundamental level, and with it most of the structure of differentiable Riemannian manifolds. Instead, spacetime manifolds and  their geometry are expected to emerge from a discrete regime based on causal structure, at scales which are large compared to the discreteness scale.\\

To reconstruct a manifold from a manifold-like causal set\footnote{A typical causal set is not manifold-like, in the sense that it does not arise with high probability from the Poisson distribution and/or that it cannot be embedded into any manifold with curvature scales much bigger then the discreteness scale.} is  in general an unsolved problem. For the other direction, namely  to construct a causal set from a given manifold,  one uses the Poisson distribution Eq.~\eqref{eq:Poisson}. One ``sprinkles" spacetime points into the manifold according to the Poisson distribution, then constructs the links (i.e., the irreducible causal relations  $x \preceq y$, which satisfy that $|\{z \in C, x\preceq z\preceq y\}|=0$) 
and then ``forgets" about the manifold and any additional structure associated with it (e.g., the tangent spaces attached to each point in the manifold). Such sprinklings will be crucial both in deriving the continuum limit of the operators we will consider, and also in performing numerical simulations of these operators.

\section{The curvature squared operator and its continuum limit} \label{sec:constructionR2}
On a manifold with boundaries, there are four different operators with mass-dimension four, i.e., constructed with four powers of derivatives. These are
\begin{equation}
	R^2,\,\quad R_{\mu\nu}R^{\mu\nu},\, \quad R_{\mu\nu\kappa\lambda}R^{\mu\nu\kappa\lambda},\,\quad \Box R. 
\end{equation}
The Riemann-invariant can also be traded for the square of the Weyl tensor; and in four dimensions a linear combination of the first three becomes a topological invariant through the Gauss-Bonnet theorem.\\
In general, a basis can be set up using any four linear combinations of the above operators. As we will find,  we can construct $R^2 - 2 \Box R$  by generalizing the operator $B$ constructed in \cite{Sorkin:2007qi,Benincasa:2010ac,Benincasa:2010as,Dowker:2013vba,Belenchia:2015hca}.\\

The scalar curvature of a causal set can be extracted from the operator $B$ \cite{Sorkin:2007qi,Benincasa:2010ac}, which is a discrete operator that converges to $\left(\Box - \frac{1}{2}R\right)$ in the continuum limit, i.e., 
\begin{eqnarray}
	\underset{\ell \rightarrow 0}{\rm lim} \, B(\phi(x)) = \left(\Box - \frac{1}{2}R(x) \right)\phi(x),
\end{eqnarray}
where  $\ell$ is the discreteness scale and $\phi(x)$ is a scalar field (which should be twice differentiable and finite within a region of compact support). By setting $\phi(x)$ to a constant ($\phi(x) = -2$ within a compact region including a neighborhood of $x$), we extract the scalar curvature:
\begin{align}
\label{eq:RicciScalar_CS}
	\underset{\ell\rightarrow 0}{\rm lim} \,B (-2)  = R(x).
\end{align}

Based on Eq.~\eqref{eq:RicciScalar_CS}, we conjecture that higher orders in the Ricci scalar can be obtained as follows: 
Given that $\psi(x)=B \phi(x) $ is just another scalar field, we can apply the discrete operator $B$ to $\psi(x)$ and take the continuum limit:
\begin{eqnarray}
	\underset{\ell\rightarrow 0}{\rm lim} \,B \left( B\, \phi(x)\right)
	= \left(\Box - \frac{1}{2}R(x) \right) \! \left(\Box - \frac{1}{2}R(x) \right) \! \phi(x),
\end{eqnarray}
Setting $\phi(x)$ to a constant value ($\phi(x) = 4$), we find
\begin{align}\label{eq:Bsq_expected}
	\underset{\ell \rightarrow 0}{\rm lim} \, B \left( B (4) \right) = R(x)^2 - 2\, \Box R(x) .
\end{align}
For constant-curvature manifolds, this expression directly gives $R^2$. On a general manifold, we obtain a combination of $R^2$ and $\Box R$. If $B^2$ is used to construct an action, it reduces to $R^2$ on manifolds without boundaries, where $\Box R$ is a total derivative that gives a vanishing contribution.

To show that Eq.~\eqref{eq:Bsq_expected} holds, we first review how to take the continuum limit of $B$ and then generalize to $B^2$.

\subsection{Review of the $B$-operator in causal sets}
 A discrete derivative operator  contains the sum over field values at nearest neighbors, next-to-nearest-neighbors etc., of a given point. This rationale underlies the construction of $B$ in causal sets \cite{Sorkin:2007qi}. There, the nearest neighbors are elements with a causal connection and no other element inbetween; the next-to-nearest-neighbors are causally connected elements with one element inbetween and so on. Here, it is not the graph-distance that determines the neighbor-relations, but instead a causal notion of distance.
 In order to build either a retarded (advanced) operator, only causally related elements to the past (future) of a given point are included.  We therefore need the layers $L_i(x)$, which contain points to the past (future) of a point which have a fixed number of points inbetween themselves and $x$. The first layer, $L_1(x)$, contains all elements that share a link with the element at $x$. The $i$-th layer is defined using $x\prec y$ to denote that $x$ precedes $y$ and is not equal to $y$.
\begin{eqnarray}
L_i(x) = \{y \in C, |\{z \in C, y \prec z \prec x\}|=i-1\},
\end{eqnarray}
i.e., the $i$-th layer includes points $y$ in the past of $x$ with $i-1$ points lying in the causal interval between $x$ and $y$. Accordingly, the second layer includes all causal intervals which are chains containing three elements in total (the top and bottom element and one intermediate one). In contrast, three-element chains which are not causal intervals are not included.

Thus,  the discrete derivative-squared operator $B$ is defined by its action on a scalar field $\phi(x)$ according to
\begin{eqnarray}\label{eq:B-Operator_Def}
	B(\phi(x)) = \frac{1}{\ell^2} \left( \alpha_d \, \phi(x) + \beta_d \, \sum_{i=1}^{n_d} C_i^{(d)}  \sum_{y \in L_i(x)} \phi(y) \,\right) \,.
\end{eqnarray}
 The prefactor $1/\ell^2$ is there to account for the mass-dimension of $B$; $\ell^2 B$ is a dimensionless quantity in any number of spacetime dimensions $d$.

 \begin{figure}[!t]
 \centering
 \includegraphics[width = 0.8\linewidth]{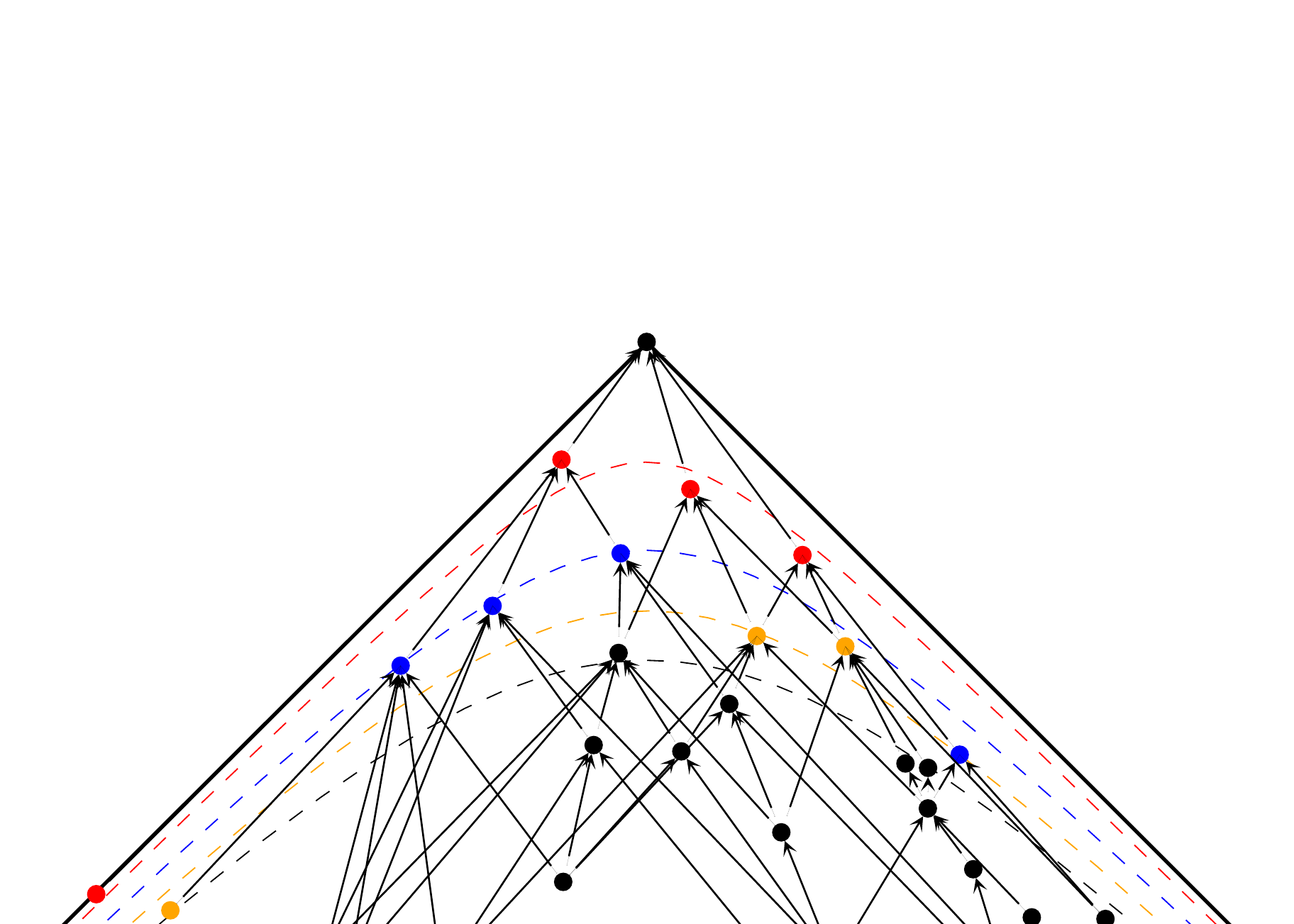}
 \caption{\label{fig:layers} Example of a causal set sprinkled into a cone in two-dimensional Minkowski space. Elements in the first layer of the element at the tip of the cone are colored red, elements in the second layer are colored blue and elements in the third layer are colored orange. The expectation values for the locations of the boundaries between layers are indicated by dashed lines.}
 \end{figure}
 
The parameters $\alpha_d$, $\beta_d$, $n_d$ and $C_i^{(d)}$, see Tab.~\ref{tab:parameters} are chosen in such a way that $\lim_{\ell
\to0}B(\phi(x))$ converges to $\left( \Box - \frac{1}{2} R(x) \right)\!\phi(x)$ for a causal set generated by  sprinkling points into a $d$-dimensional manifold \cite{Benincasa:2010ac,Dowker:2013vba,Glaser:2013xha}. 

\begin{table}
\begin{center}
\begin{tabular}{|c|c|c|c|c|c|c|c|}
\hline
$d$& $\alpha_d$ & $\beta_d$ & $n_d$ & $C_1^{(d)}$& $C_2^{(d)}$& $C_3^{(d)}$ & $C_4^{(d)}$\\ \hline \hline
2 & -2 & 4 & 3 & 1 & -2 & 1 & -\\ \hline\hline
4 & $\frac{-4}{\sqrt{6}}$ & $\frac{4}{\sqrt{6}}$& 4 & 1 & -9 & 16 & -8.\\\hline
\end{tabular}
\end{center}
\caption{\label{tab:parameters} We show the parameters that enter Eq.~\eqref{eq:B-Operator_Def} for $d=2$ (see \cite{Sorkin:2007qi}) and $d=4$ (see \cite{Benincasa:2010ac}); for the $d$-dimensional generalization, see \cite{Dowker:2013vba,Glaser:2013xha}.}
\end{table}

For causal sets obtained by a Poisson sprinkling into a $d$-dimensional Lorentzian manifold $(\mathcal{M},g)$,  with metric $g_{\mu\nu}$, with sprinkling density $\rho=
 \ell
^{-d}$, the expectation value of the random variable $B(\phi(x))$ can be written as
\begin{eqnarray}
	\langle B(\phi(x))\rangle & =& \frac{1}{\ell^2} \bigg(  \alpha_d \, \phi(x)  \\
	& &+ \frac{\beta_d}{\ell^d}\, \sum_{i=1}^{n_d}  C_i^{(d)}  
	\!\int_{J^-(x)}d^d y \,\sqrt{-g(y)} \,\, \frac{\big(\ell^{-d} \,\mathcal{V}(x,y) \big)^{i-1} }{(i-1)!}e^{\ell^{-d} \, \mathcal{V}(x,y)} 
	\, \phi(y) \bigg)\,, \nonumber
\end{eqnarray}
where $J^-(x)$ is the causal past of $x$ and $\mathcal{V}(x,y)$ is the volume of the causal interval between points $x$ and $y$, see App.~\ref{app:ExpctValue} for more details.

For compactness, we summarize the summation over past layers as
\begin{eqnarray}\label{eq:Integral_I_Def}
	\mathcal{I}_d[\phi](x) = \frac{1}{{\ell}^d}\,  \int_{J^-(x)} \hspace*{-.6cm} d^d y \,\sqrt{-g(y)} \,\, \mathcal{P}\big( {\ell}^{-d} \,\mathcal{V}(x,y) \big)\, \phi(y) \,,
\end{eqnarray} 
with
\begin{eqnarray}
	\mathcal{P}\big( l^{-d} \,\mathcal{V}(x,y) \big) = \sum_{i=1}^{n_d}  C_i^{(d)} \frac{\big({\ell}^{-d} \,\mathcal{V}(x,y) \big)^{i-1} }{(i-1)!}\,e^{-{\ell}^{-d} \, \mathcal{V}(x,y)} \,.
\end{eqnarray}
With this notation, we can rewrite $\langle B(\phi(x)) \rangle$ as
\begin{align}\label{eq:B_operator_ShortNotation}
	\langle B(\phi(x))\rangle = {\ell}^{-2}\, \big( \alpha_d  \, \phi(x) + \beta_d \,\mathcal{I}_d[\phi](x) \big)  \,.
\end{align}

In the continuum limit, $\ell \to 0$, we can compute $\mathcal{I}_d[\phi](x)$ by following \cite{Belenchia:2015hca}. 
 In summary, to compute the continuum limit of $\mathcal{I}_d[\phi](x)$ the authors of \cite{Belenchia:2015hca} decompose the integration region $J^-(x)$ into three sub-regions $J^-(x) = W_1 \cup W_2 \cup W_3$. 
$W_2$ and $W_3$ are composed of space-time points that are far away from $x$, and their contributions to $\mathcal{I}_d[\phi](x)$ are sub-leading in the continuum limit $\ell \to0$.
The leading contribution to $\mathcal{I}_d[\phi](x)$ comes from an integration region $W_1 \subset J^-(x)$, which includes only points that are sufficiently close to $x$. Within the region $W_1$, the authors of \cite{Belenchia:2015hca}: i) expand $\phi(y)$ in a Taylor series around $x$ and ii) expand the metric $g_{\mu \nu}$ and the causal volume $\mathcal{V}(x,y)$ in terms of Riemann normal coordinates around $x$. \\
 For a detailed explanation of the steps involved in the calculation of $\mathcal{I}_d[\phi](x)$ see, e.g., \cite{Belenchia:2015hca,Dowker:2013vba,Glaser:2013xha,Christopherthesis}.\\
Computing $\mathcal{I}_d[\phi](x)$ in a power series in the discreteness scale $\ell$, we find
\begin{eqnarray}\label{eq:Integral_I_Sol}
	\mathcal{I}_d[\phi](x) = - \beta_d^{-1}\left( \alpha_d  - \ell^2  \,\Delta \right)\! \phi(x) + {\ell}^4 \, \Omega_\phi(x) \, .
\end{eqnarray}
Here $\Delta  =  \Box  - \frac{1}{2}\, R(x) $
and $\Omega_\phi(x)$ contains possible higher order contributions, such that $\ell^2\,\Omega_\phi(x) \to 0$ when $\ell \to 0$. 

From Eqs.~\eqref{eq:Integral_I_Sol} and \eqref{eq:B_operator_ShortNotation}, we find
\begin{eqnarray}
	\langle B(\phi(x)) \rangle = \left( \Box  - \frac{1}{2}\, R(x) \right)\! \phi(x)
	+\beta_d\,\ell^2\, \Omega_\phi(x) \,.
\end{eqnarray}
Thus, $\langle B(\phi(x)) \rangle$ converges to $\left( \Box  - \frac{1}{2}\, R(x) \right)\!\phi(x)$ in the continuum limit ($\ell\to 0$).
In the continuum limit of $\langle B(\phi(x))\rangle$, the relevant terms come from contributions up to second order in the Taylor expansion of $\phi(x)$  and up to first order in the expansion in Riemann normal coordinates. The higher-order terms included in $\Omega_\phi(x)$ are suppressed by positive powers of the discreteness scale $\ell$.

\subsection{$B^2$-operator and its continuum limit}

We define the discrete operator $B^2$ by the successive application of $B$ on a scalar field $\phi(x)$, namely
\begin{eqnarray}
	B^2(\phi(x)) \coloneqq B(B(\phi(x))) \,.
\end{eqnarray} 
We will next derive the continuum limit of the expectation value of this operator, to show that it produces the expected higher-derivative terms.
 As a first step, we write $B(\phi(x))$ using the result in Eq.~\eqref{eq:B-Operator_Def}. The resulting expression contains $B(\phi(x))$ and $\sum_{y \in L_i(x)} B(\phi(y))$, both of which can again be written using Eq.~\eqref{eq:B-Operator_Def}. Thus, we find
\begin{eqnarray}
	B^2(\phi(x)) &=& \frac{1}{{\ell}^2} \Bigg( \alpha_d \, B(\phi(x)) + \beta_d \, \sum_{i=1}^{n_d} C_i^{(d)} \!\!\sum_{y \in L_i(x)} B(\phi(y)) \,\Bigg) ,  \nonumber \\
	&=& \frac{1}{{\ell}^4} \Bigg( \alpha_d^2 \,\phi(x) + 2 \,\alpha_d \beta_d \, \sum_{i=1}^{n_d} C_i^{(d)} \!\!\sum_{y \in L_i(x)} \phi(y)  \nonumber\\
	&{}&\,\,\quad+ \beta_d^2 \sum_{i,j=1}^{n_d} C_i^{(d)} C_j^{(d)} \!\!\sum_{y \in L_i(x)} \sum_{z \in L_j(y)} \phi(z) \Bigg) .  \label{eq:Bsq_extendend}
\end{eqnarray}

For causal sets obtained by a sprinkling process, the expectation value of the random variable $B^2(\phi(x))$ can be written as (see the appendix for details)
\begin{eqnarray}
	&{}&\langle B^2(\phi(x)) \rangle 
	= \frac{1}{{\ell}
		^4} \Bigg( \alpha_d^2 \, \phi(x) + \frac{2\,\alpha_d \beta_d}{{\ell}^d}\, \!\int_{J^-(x)} \hspace*{-.6cm} d^d y \,\sqrt{-g(y)} \,\, \mathcal{P}\big( { \ell}^{-d} \,\mathcal{V}(x,y) \big) \, \phi(y)  \label{eq:Bsqexp}  \\
	&{}&
	\qquad\qquad +\frac{\beta_d^2}{{\ell}^{2d}} \,  \!\int_{J^-(x)} \hspace*{-.6cm} d^d y \,\sqrt{-g(y)}  
	\int_{J^-(y)} \hspace*{-.6cm} d^d z \,\sqrt{-g(z)} \,\,\mathcal{P}\big({\ell}^{-d} \,\mathcal{V}(x,y) \big) \,\mathcal{P}\big({\ell}^{-d} \,\mathcal{V}(y,z) \big) \, \phi(z) \Bigg) .\nonumber
\end{eqnarray}
 In this expression, $\ell$ occurs with powers that depend on $d$, but also as a $1/\ell^4$ prefactor. This prefactor carries the mass-dimension of $B^2$ and is therefore not dimension-dependent.

Eq.~\eqref{eq:Bsqexp} can be rewritten in terms of $\Box \phi$, $R\, \phi$ and $\Omega_{\phi}$ using Eq.~\eqref{eq:Integral_I_Sol}.
Since $\langle B^2(\phi(x)) \rangle$ contains an overall factor ${\ell}^{-4}$, we might expect that the term ${\ell}^4 \, \Omega_\phi(x)$ in the second term in Eq.~\eqref{eq:Bsqexp} generates a finite contribution in the continuum limit (${\ell}\to0$). Such a contribution would require the inclusion of higher-order terms in the Taylor-expansion of $\phi(x)$ and higher-order terms in the expansion  in Riemann normal coordinates. 
However, this ${\ell}^4 \, \Omega_\phi(x)$-term cancels out with a similar contribution  in the third term of Eq.~\eqref{eq:Bsqexp}. Therefore,  taking the continuum limit $\langle B^2(\phi(x)) \rangle$ does not require the explicit evaluation of the higher-order terms contained in $\Omega_\phi(x)$.

From Eq.~\eqref{eq:Integral_I_Sol},  we find
\begin{eqnarray}
	&{}& \frac{\beta_d^2}{{\ell}^{2d}} \,  \!\int_{J^-(x)} \hspace*{-.6cm} d^d y \,\sqrt{-g(y)} \int_{J^-(y)} \hspace*{-.6cm} d^d z \,\sqrt{-g(z)} \,\,\mathcal{P}\big({\ell}^{-d} \,\mathcal{V}(x,y) \big) \,\mathcal{P}\big( {\ell}^{-d} \,\mathcal{V}(y,z) \big) \, \phi(z) \nonumber\\
	&{}&\qquad= - \frac{\alpha_d \beta_d}{\ell^d}\,  \int_{J^-(x)} \hspace*{-.6cm} d^d y \,\sqrt{-g(y)} \,\, \mathcal{P}\big({\ell}^{-d} \,\mathcal{V}(x,y) \big)\, \phi(y) \label{eq:IcompI_Sol1} \\
	&{}&\qquad\quad+ \frac{\beta_d}{\ell^{d-2}}\,  \int_{J^-(x)} \hspace*{-.6cm} d^d y \,\sqrt{-g(y)} \,\, \mathcal{P}\big({\ell}^{-d} \,\mathcal{V}(x,y) \big)\, \Delta \phi(y) \label{eq:IcompI_Sol2} \\
	&{}&\qquad\quad+ \frac{\beta_d^2}{{\ell}^{d-4}}\,  \int_{J^-(x)} \hspace*{-.6cm} d^d y \,\sqrt{-g(y)} \,\, \mathcal{P}\big({\ell}^{-d} \,\mathcal{V}(x,y) \big)\, \Omega_\phi(x) \,.\label{eq:IcompI_Sol3} 
\end{eqnarray}
The second line, \eqref{eq:IcompI_Sol1} corresponds to $\mathcal{I}[\phi](x)$, thus we can use the expression given by Eq.~\eqref{eq:Integral_I_Sol}. The integrals in the third and fourth line, \eqref{eq:IcompI_Sol2} and \eqref{eq:IcompI_Sol3} have the same structure as $\mathcal{I}[\phi](x)$ with $\phi$ replaced by $\Delta \phi$ and $\Omega_{\phi}$, respectively. 
 We can always substitute $\Delta \phi$ and $\Omega_{\phi}$ by some scalar field $\psi(x)$, for which we use the prescription in Eq.~\eqref{eq:Integral_I_Sol}.
Therefore, we can compute the third and fourth line  from Eq.~ \eqref{eq:Integral_I_Sol}. The resulting expressions are
\begin{eqnarray}
	\hspace*{-.5cm}\frac{1}{\ell^d}\int_{J^-(x)} \hspace*{-.6cm} 
	d^d y \,\sqrt{-g(y)} \,\, \mathcal{P}\big( \ell^{-d} \,\mathcal{V}(x,y) \big)\, \Delta \phi(y) 
	&=&-\beta_d^{-1}\left( \alpha_d - \ell^2 \Delta \right)  \Delta  \phi(x) + {\ell}^4 \Omega_{\Delta\phi}(x)\,,\\\hspace*{-.5cm}\frac{1}{\ell^d}\int_{J^-(x)} \hspace*{-.6cm} d^d y \,\sqrt{-g(y)} \,\, \mathcal{P}\big({\ell}^{-d} \,\mathcal{V}(x,y) \big)\, \Omega_\phi(y) 
	&=&-\beta_d^{-1}\left( \alpha_d - \ell^2 \Delta \right) \Omega_\phi(x) + {\ell}^4 \Omega_{\Omega_\phi}(x)\,.
\end{eqnarray}
Combining these results with \eqref{eq:IcompI_Sol1}, \eqref{eq:IcompI_Sol2} and \eqref{eq:IcompI_Sol3}, we find
\begin{eqnarray}\label{eq:IcompI_Sol4}
	\begin{aligned}
		&{}\frac{\beta_d^2}{{\ell}^{2d}} \,  \!\int_{J^-(x)} \hspace*{-.6cm} d^d y \,\sqrt{-g(y)}\int_{J^-(y)} \hspace*{-.6cm} d^d z \,\sqrt{-g(z)} \,\,\mathcal{P}\big({\ell}^{-d} \,\mathcal{V}(x,y) \big) \,\mathcal{P}\big({\ell}^{-d} \,\mathcal{V}(y,z) \big) \, \phi(z)  \\
		&{}\qquad=\alpha_d^2 \phi(x)  - 2\,\alpha_d \,{\ell}^2  \,\Delta \phi(x) + \ell^4 \Delta^2 \phi(x)   \\
		&{}\qquad\,\, - 2\,\alpha_d \beta_d \, \ell^4 \, \Omega_\phi(x)  +  \beta_d \ell^6 \Omega_{\Delta\phi}(x) + \beta_d\, \ell^6 \Delta \Omega_\phi(x) +\beta_d^2\,\ell^8 \Omega_{\Omega_\phi}(x) \,. 	
	\end{aligned}
\end{eqnarray}
 In here, we find terms at $\mathcal{O}(\ell^0)$ and $\mathcal{O}(\ell^2)$. These in fact cancel with similar terms from the first line of Eq.~\eqref{eq:Bsqexp}.
 We get the following result
\begin{eqnarray}
	\langle B^2(\phi(x)) \rangle = \Delta^2 \phi(x) + \beta_d \,{\ell}^2  \big( \Delta \Omega_\phi(x) + \Omega_{\Delta\phi}(x) \big) + {\mathcal{O}(\ell^4)}.
\end{eqnarray}
Therefore, the continuum limit of $\langle B^2(\phi(x)) \rangle$ converges to
\begin{eqnarray}
	\lim_{{\ell}\to 0} \langle B^2(\phi(x)) \rangle 
	&=& \left( \Box - \frac{1}{2} R(x) \right)\!\left( \Box - \frac{1}{2} R(x) \right)\!\phi(x) \, , \nn \\
	&=& \Box^2 \phi(x) - \frac{1}{2}R(x) \Box \phi(x) - \frac{1}{2} \Box \big( R(x) \,\phi(x) \big) + \frac{1}{4} R(x)^2 \, \phi(x)\,.
\end{eqnarray}

\subsection{Disentangling $R(x)^2$ and $\Box R(x)$}

The continuum limit of $\langle B^2(\phi(x)) \rangle$ leads to a combination of $R(x)^2$ and $\Box R(x)$. For a constant scalar field, $\phi(x)=4$, we find
\begin{eqnarray}\label{eq:B2_const_phi}
	\lim_{\ell\to0} \langle B^2(4) \rangle  =R(x)^2 - 2  \,\Box R(x) \,. 
\end{eqnarray}
To disentangle $R(x)^2$ and $\Box R(x)$, we consider a second object, namely the square of $\langle B(\phi(x)) \rangle$, which allows to extract $R(x)^2$ independently of $\langle B^2(\phi(x)) \rangle$. From Eq.~\eqref{eq:RicciScalar_CS}, we observe that
\begin{eqnarray}\label{eq:B_const_phi_square}
	\lim_{\ell\to0} \langle B(-2) \rangle^2 = R(x)^2   \,.
\end{eqnarray}
Combining Eqs.~\eqref{eq:B2_const_phi} and \eqref{eq:B_const_phi_square}, we get
\begin{align}
	\Box R(x) = \frac{1}{2} \lim_{\ell\to0} \left( \langle B(-2) \rangle^2 - \langle B^2(4) \rangle  \right)  \,.
\end{align}

The two expressions $\langle B^2\rangle$ and $\langle B \rangle^2$ both come with challenges when it comes to their numerical evaluation.\\
 It was shown that $\langle B \rangle$ fluctuates quite significantly; therefore we expect that $\langle B \rangle^2$ can also have rather large fluctuations. This may require large simulations for the mean value to converge sufficiently.\\
 $\langle B^2\rangle$ was not studied before, but the numerical challenge is clear: whereas $\langle B \rangle$ requires information on four layers into the past of a point (in four spacetime dimensions), $\langle B^2\rangle$ requires information on eight layers into the past. This requires large enough causal sets, such that boundary effects can be avoided in the numerical study. In preliminary numerical studies, we find that causal sets which are large enough so that $\underset{\ell \rightarrow 0}{\rm lim}\, \langle B \rangle$ converges to the expected value, are not large enough so that $\underset{\ell \rightarrow 0}{\rm lim}\, \langle B^2 \rangle$ converges as well. A systematic study of the volume scaling of $\underset{\ell \rightarrow 0}{\rm lim}\, \langle B^2 \rangle$ is therefore called for, in particular for the so-called ``smeared" operator, where the smearing dampens fluctuations \cite{Benincasa:2010as,Dowker:2013vba,Belenchia:2015hca}.

\section{Generalization to higher orders} \label{sec:higher_order_operators}

We can generalize our construction of the $B^2$-operator to higher orders. 
We define recursively
\begin{eqnarray}
	B^{n}(\phi(x)) \coloneqq B(B^{n-1}(\phi(x)))  \,, \qquad (n\geq 2)\,,
\end{eqnarray}
with $n=1$ being the standard $B$-operator given by \eqref{eq:B-Operator_Def}. From the recursive definition of $B^{n}(\phi(x))$, we can show that
\begin{eqnarray}
	B^{n}(\phi(x)) = \ell^{-2n} \left( \alpha_d^n \phi(x) + \sum_{k=1}^{n} \binom{n}{k} \alpha_d^{n-k} \, \beta_d^k \,b_k[\phi(x)] \right) \,,
\end{eqnarray}
with $b_k$ defined as
\begin{align}
	b_k[\phi(x)] = \sum_{i_1,\cdots,i_k=1}^{n_d} C_{i_1} \cdots C_{i_k} 
	\sum_{y_1 \in L_{i_1}(x)} \sum_{y_2 \in L_{i_2}(y_1)} \cdots  \sum_{y_k \in L_{i_k}(y_{k-1})}
	\phi(y_k) \,.
\end{align}

For causal sets obtained by sprinkling process, we can show that the expectation value of the random variable $B^n(\phi(x))$ satisfies the recursive relation, see the appendix for details, 
\begin{equation}\label{eq:Bn-expect_value_recursive}
	\begin{aligned}
		\langle B^{n}(\phi(x)) \rangle &=
		\ell^{-2} \Bigg( \alpha_d \, \langle B^{n-1}(\phi(x)) \rangle  \\
		&+ \frac{\beta_d}{\ell^d}  \int_{J^{-}(x)} \hspace*{-.6cm} d^dy \sqrt{-g(y)} 
		\,  \mathcal{P}\left( \ell^{-d} \mathcal{V}(x,y) \right)  \, \langle B^{n-1}(\phi(y)) \rangle \Bigg) .
	\end{aligned}
\end{equation}
We now show by induction that this recursive relation for $\langle B^{n}(\phi(x)) \rangle$ implies the following behavior
\begin{align}\label{eq:Bn-expect_value}
	\langle B^n(\phi(x)) \rangle = 
	\Delta^n\phi(x) + \beta_d \, \ell^2 \, \Omega_{B^{n-1}(\phi)}(x) \,.
\end{align}
In Sec.~\ref{sec:constructionR2}, we showed that this formula is valid for $n=1$ and $n=2$. To complete the proof by induction we need to show that \eqref{eq:Bn-expect_value} implies a similar relation for $n+1$. Using \eqref{eq:Bn-expect_value_recursive} with $n\to n+1$, we find
\begin{equation}\label{eq:Bnplus1_intermed}
	\begin{aligned}
		\langle B^{n+1}(\phi(x)) \rangle &=
		\ell^{-2} \Bigg( \alpha_d \, \langle B^{n}(\phi(x)) \rangle  \\
		& + \frac{\beta_d}{\ell^d}  \int_{J^{-}(x)} \hspace*{-.6cm} d^dy \sqrt{-g(y)} 
		\, \mathcal{P}\left( \ell^{-d} \mathcal{V}(x,y) \right) \langle B^{n}(\phi(y)) \rangle \Bigg)  ,\\
	    &=\ell^{-2} \Bigg( \alpha_d \,\left( \Delta^n \phi(x) +
		\beta_d \, \ell^2 \, \Omega_{B^{n-1}(\phi)}(x) \right)  \\  
		&\hspace*{1cm}+ \frac{\beta_d}{\ell^d}  \int_{J^{-}(x)} \hspace*{-.6cm} d^dy \sqrt{-g(y)} 
		\, \mathcal{P}\left( \ell^{-d} \mathcal{V}(x,y) \right)  \, \Delta^n \phi(y)  \\ 
		&\hspace*{1cm}+ \frac{\beta_d^2}{\ell^{d-2}} \int_{J^{-}(x)} \hspace*{-.6cm} d^dy \sqrt{-g(y)} 
		\,\mathcal{P}\left( \ell^{-d} \mathcal{V}(x,y) \right) \,\Omega_{B^{n-1}(\phi)}(y)  \Bigg) .
	\end{aligned}
\end{equation}
To evaluate the integrals in the last two lines we can use a similar formula as in \eqref{eq:Integral_I_Sol}, which results in
\begin{eqnarray}
	&{}&\frac{1}{\ell^d}\int_{J^-(x)} \hspace*{-.6cm} d^d y \,\sqrt{-g(y)} \,\, \mathcal{P}\left( \ell^{-d} \,\mathcal{V}(x,y) \right)\, \Delta^n \phi(y) \nonumber \\
	&{}&\qquad= - \beta_d^{-1}\left( \alpha_d - \ell^2 \Delta \right)  \Delta^n  \phi(x) + \ell^4 \Omega_{\Delta^n\phi}(x)\,,\\
	&{}&\frac{1}{\ell^d}\int_{J^-(x)} \hspace*{-.6cm} d^d y \,\sqrt{-g(y)} \,\, \mathcal{P}\big( \ell^{-d} \,\mathcal{V}(x,y) \big)\, \Omega_{B^{n-1}(\phi)}(y)  \nonumber\\
	&{}&\qquad= - \beta_d^{-1}\left( \alpha_d - \ell^2 \Delta \right) \Omega_{B^{n-1}(\phi)}(y) + \ell^4\Omega_{\Omega_{B^{n-1}(\phi)}}(x)\,.
\end{eqnarray}
Plugging these integrals back into \eqref{eq:Bnplus1_intermed}, we find
\begin{align}\label{eq:Bnplus1-expect_value}
	\langle B^{n+1}(\phi(x)) \rangle = 
	\Delta^{n+1}\phi(x) + \beta_d \, \ell^2 \, \Omega_{B^{n}(\phi)}(x) \,,
\end{align}
with
\begin{align}
	\Omega_{B^{n}(\phi)}(x) =  \Omega_{\Delta^n\phi}(x) + \Delta \Omega_{B^{n-1}(\phi)}(x) +
	\ell^2\,\Omega_{\Omega_{B^{n-1}(\phi)}}(x) \,,
\end{align}
which completes our proof by induction.\\

From Eq.~\eqref{eq:Bn-expect_value}, we can see that $\langle B^n (\phi(x))\rangle$ has a continuum limit given by
\begin{eqnarray}\
	\lim_{\ell\to 0} \langle B^n(\phi(x)) \rangle = \left( \Box - \frac{1}{2} R(x) \right)^{\!\! n}\!\phi(x)\,.
\end{eqnarray}
To extract information on $R(x)^n$ we can set $\phi(x)$ to a constant value, which leads to 
\begin{align}\label{eq:R^n}
	R(x)^n= 
	\, \lim_{\ell\to0} \, \langle B^n((-2)^n) \rangle  
	- (-2)^{n-1} \,\Box^{n-1}R(x) \,+ \cdots \,,
\end{align}
where the dots indicate a sequence of terms constructed with combinations of $\Box$ and $R(x)$. 

Similarly to the case of $B^2$, our construction of $B^n$ provides the superposition $R(x)^n$ with terms involving $\Box R(x)$. For causal sets generated via sprinkling on manifolds with constant curvature scalar, only the first term on the rhs of \eqref{eq:R^n} survives, thus, we find $R(x)^n = \lim_{\ell\to0} \, \langle B^n((-2)^n) \rangle$. For causal sets generated via sprinkling over a manifold with non-constant $R(x)$, we can use lower order operators to disentangle the different contributions.

\section{Causal-set action with higher-derivative operators}\label{sec:action}
The discrete operator $B(\phi(x))$ allows us to define the Benincasa-Dowker action for causal sets \cite{Benincasa:2010ac}, which is a discrete version of the Einstein-Hilbert action.
 We can generalize the Benincasa-Dowker action to include higher-curvature terms, e.g., by including the $B^2$-operator,
\begin{eqnarray}\label{eq:discreteAction}
	S[C] = \sum_{x \in C} \ell^4 \left(\frac{1}{16\pi G_\textmd{N}} B(-2) + b^2\,B^2(4)  \right).
\end{eqnarray}
Here, we have introduced two couplings, namely the Newton coupling in front of the curvature term, and a coupling $b^2$ in front of the curvature-squared term. 

Based on the results of \ref{sec:constructionR2}, the action \eqref{eq:discreteAction} can be viewed as a discretization of 
\begin{align}
	S_{\rm higher\, order} = \int d^d x \, \sqrt{-g} \left(\frac{1}{16\pi G_\textmd{N}} R + b^2\,R^2 - 2\,b^2\,\Box R \right) + \mathcal{O}(\ell^2).
\end{align}
At the level of the action, the contribution $\Box R$ in the higher-derivative term becomes a total derivative. If we assume a manifold without boundaries, the action therefore only contains the terms $R$ and $R^2$. On manifolds with boundaries, an additional boundary contribution $\sim \Box R$ is present.\\

\begin{figure}[!t]
\begin{center}
\includegraphics[width=\linewidth, clip=true, trim=15cm 2cm 0cm 0.5cm]{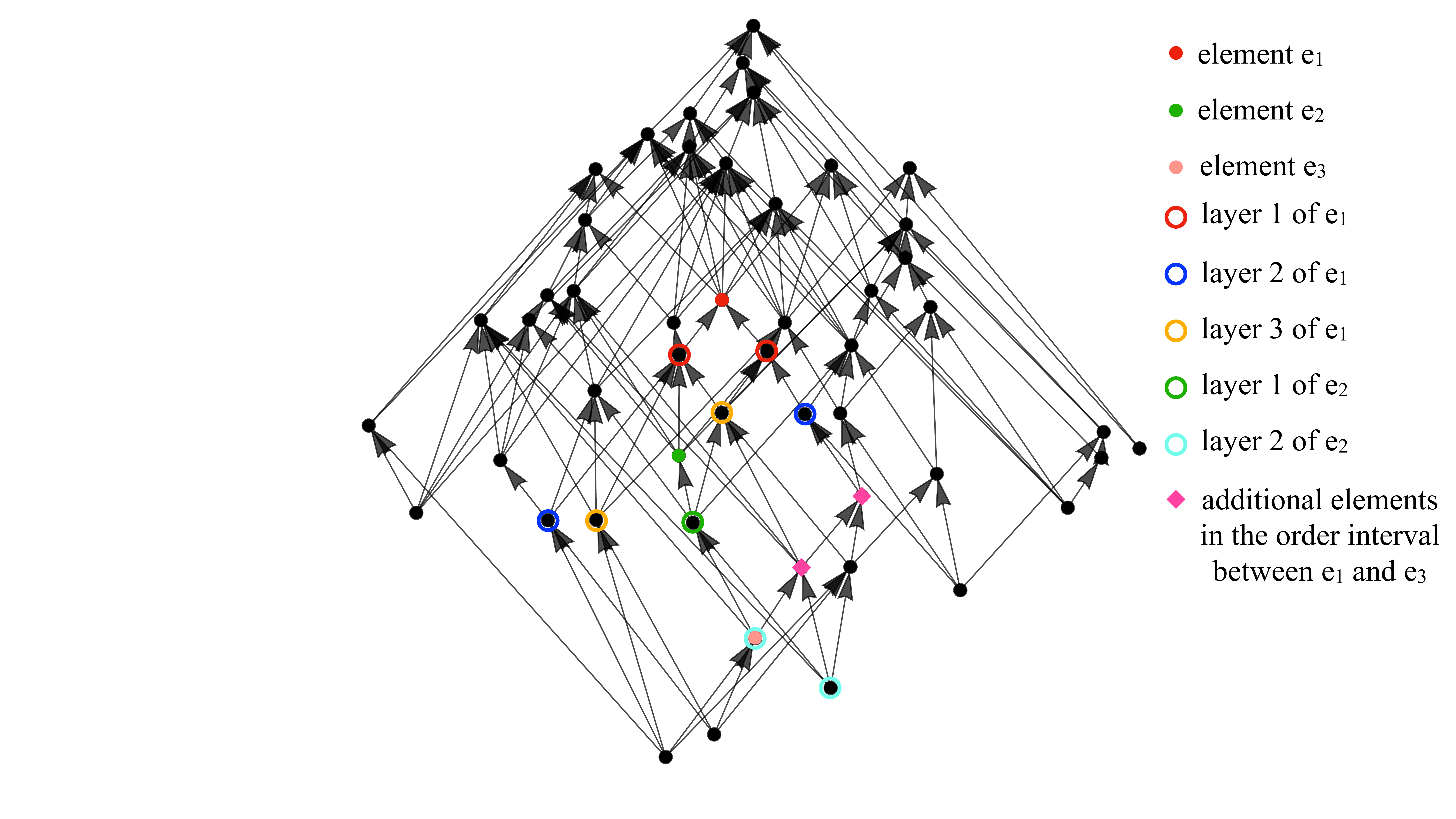}
\end{center}
\caption{\label{fig:stackedorders} We show a sprinkling into Minkowski spacetime which illustrates stacked order intervals: The green causal-set element $e_2$ lies in layer 4 of the red causal-set element $e_1$, i.e., there are three other elements inbetween. Two layers to the past of $e_2$ lies $e_3$. The order interval between $e_1$ and $e_3$ contains additional causal-set elements (magenta diamonds) and $e_3$ does not lie in layer 6 (2+4) of $e_1$, but instead in layer 9 (with 8 elements contained in the order interval).}
\end{figure}

\begin{figure}[!t]
\begin{center}
\includegraphics[width=\linewidth]{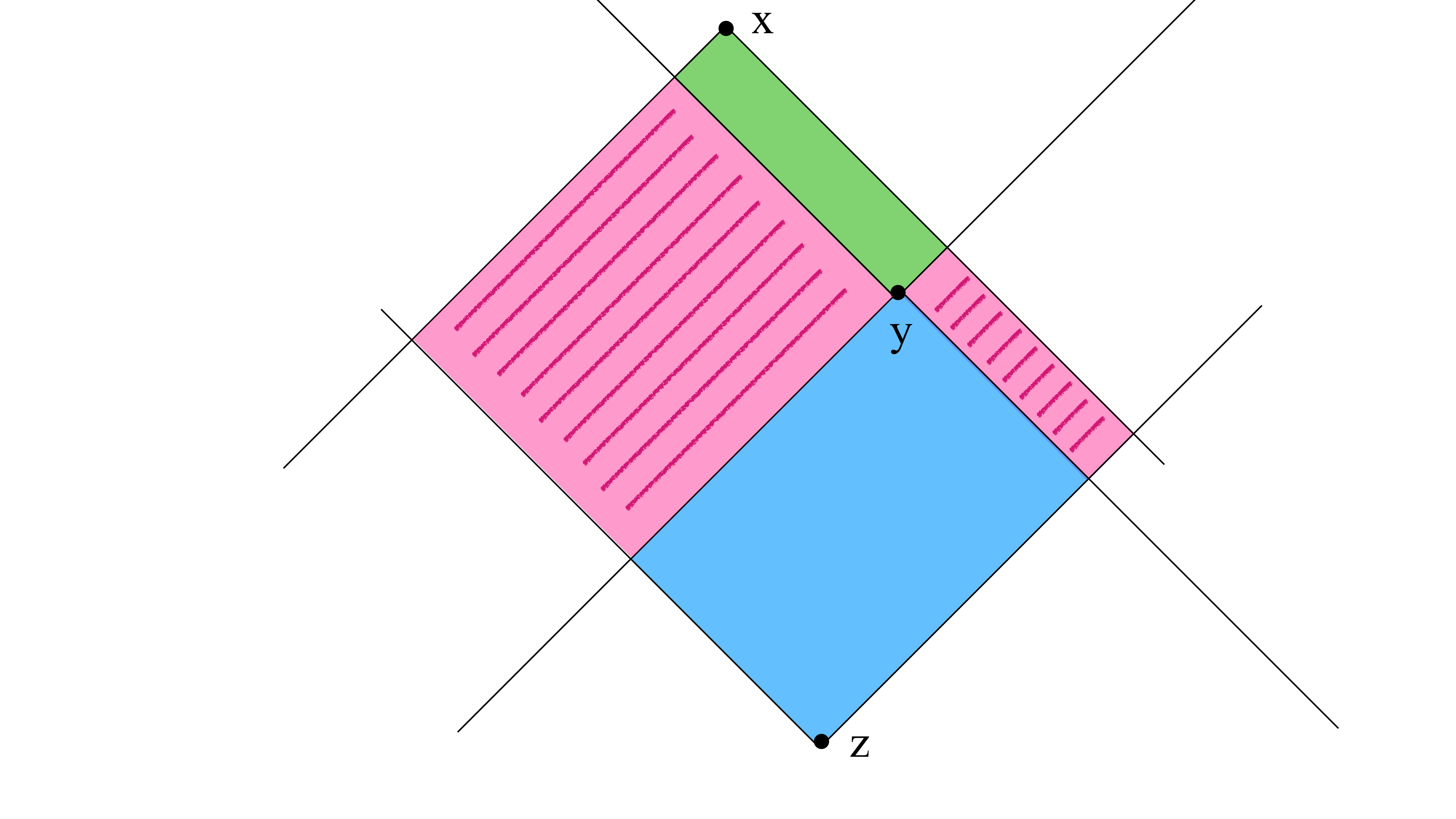}
\end{center}
\caption{\label{fig:stackedorderscont} We show a stacked order interval in two-dimensional Minkowski spacetime. The green and blue regions are part of the order interval between $x,y$ and $z$, but the magenta regions with broad lines, which are part of the order interval between $x$ and $z$, are not part of the stacked order interval.}
\end{figure}

The Benincasa-Dowker action with $b=0$ can be rewritten in terms of the number of order intervals in a causal set. An inclusive order interval $I_n$ is the causal interval between two points $i$ and $j$ of cardinality $n$; 
\begin{eqnarray}
	I_n(x,y) = \{z \in C\vert y \preceq z \preceq x, |I_n(x,y)|=n-2\}.
\end{eqnarray}
Thus, order intervals make up the layers in the definition of the d'Alembertian. We define, for a causal set $C$
\begin{eqnarray}
	N&=& \mbox{number of elements in }C,\\
	N_1&=&\mbox{number of links, i.e., number of $I_0$, in }C,\\
	N_i&=&\mbox{number of $I_{i-1}$ in }C.
\end{eqnarray}
The causal set action in four dimensions is \cite{Benincasa:2010ac}
\begin{equation}
	\begin{aligned}
		\hspace*{-.15cm} S[\mathcal{C}] &= \frac{\ell^4}{16\pi G_\textmd{N}} \sum_{x \in C}B(-2) \\
		\hspace*{-.15cm}&= \frac{\ell^2}{16\pi G_\textmd{N}}\sum_{x \in C} \Bigg(- \frac{4}{\sqrt{6}}(-2)  \\
		\hspace*{-.15cm}&\quad +  \frac{4}{\sqrt{6}}\bigg(\sum_{y \in L_1(x)}(-2)-9 \sum_{y \in L_2(x)}(-2)+ 16 \sum_{y\in L_3(x)}(-2) - 8 \sum_{y \in L_4(x)} (-2) \bigg) \Bigg) \\
		\hspace*{-.15cm}&= N -N_1+9 N_2-16 N_3+8 N_4 \,,
	\end{aligned}
\end{equation}
where we have absorbed numerical factors of order 1 in the ratio of discreteness scale and Planck scale, $\ell_\textmd{Pl.} \sim G_\textmd{N}^{1/2}$.\\
The generalized, higher-order action requires a generalization of order intervals to \emph{stacked} order intervals, because the sum $\sum_{y\in L_i(x)}\sum_{z\in L_j(y)}$ does not give rise to order intervals. This can be seen in Fig.~\ref{fig:stackedorders}, where we show that the combination of an order interval between $e_1$ and $e_2$ with an order interval between $e_2$ and $e_3$ (with $e_3 \prec e_2 \prec e_1$), is not the order interval between $e_1$ and $e_3$. We also show what stacked order intervals look like in the continuum, cf.~Fig.~\ref{fig:stackedorderscont}.

Therefore we define a stacked order interval $I_{n,m}(x,y,z)$ as the union of two order intervals, for which the point at the top of the lower interval is also the point at the bottom of the upper interval:
\begin{equation}
	\begin{aligned}
		I_{n,m}(x,y,z) &=  I_{m}(y,z) \cup I_{n}(x,y) \\
		&=  \{w \in C\vert  z \preceq w \preceq y, |I_m(y,z)|=m-2\}\,\\
		&\quad\, \cup\, \{w \in C\vert  y \preceq w \preceq x, |I_n(x,y)|=n-2\}\, .
	\end{aligned}
\end{equation}
With this definition, the higher-order action can be written as
\begin{equation}
	\begin{aligned}
		S_{\rm higher\, order}[C] &= \left(N -N_1+9 N_2-16 N_3+8 N_4 \right)\\
		&\,\,+ \tilde{b}^2 \Big[ \, N - 2\,\left( N_1- 9 N_2+ 16 N_3 - 8N_4\right) \\
		&\,\frac{}{}+  N_{1,1}- 9 N_{1,2}+ 16N_{1,3}- 8 N_{1,4} \\
		&\,\frac{}{}- 9N_{2,1} + 81 N_{2,2}-144N_{2,3}+72N_{2,4} \\
		&\,\frac{}{}+ 16N_{3,1}-144N_{3,2}+256N_{3,3}-128 N_{3,4}\\
		&\,\frac{}{}- 8N_{4,1}+72 N_{4,2}-128 N_{4,3}+64 N_{4,4} \,
		\Big] \,,
	\end{aligned}
\end{equation}
where $b^2$ and $\tilde{b}^2$ are related by a factor $\ell^4$ and numerical factors of order one. Herein, $N_{n,m}$ counts the number of stacked order intervals $I_{n,m}$. We note that $N_{n,m}\neq N_{m,n}$ for $n\neq m$, because the causal ordering of the three points matters.\\
This provides a way to study higher-order dynamics in causal sets.
\section{Conclusions and outlook}\label{sec:conclusions}
In this paper, we have presented a way of extracting higher-order scalar curvature in causal sets. Our approach is based on the successive application of the discrete $B$-operator on scalar fields.

We have shown that the operator $B^2$ has a well defined continuum limit (averaged over sprinklings) which converges to a combination of $R(x)^2$ and $\Box R(x)$. We extended the construction to higher-order and, based on a induction process, we have demonstrated that continuum limit of the discrete $B^n$ is related to $R(x)^n$ and other terms involving combinations of $\Box$ and $R(x)$.

Based on our construction of higher-order operators in causal sets, we can define a discrete action. Its continuum limit contains higher-order curvature terms beyond the Einstein-Hilbert action.  The definition of such an action is motivated by a possible connection between causal sets and asymptotic safety, where causal sets are used as a regularization of the Lorentzian path integral, for which we search an asymptotically safe continuum limit. The latter is signalled by a second (or higher) order phase transition in the phase diagram of causal sets. So far, results based on the Benincasa-Dowker action indicate that the causal set phase diagram exhibits only a first-order phase transition.\\
We conjecture that a curvature-squared term could be the missing ingredient for a second-order phase transition in the causal set phase diagram. This is motivated by several studies on asymptotically safe gravity based on functional renormalization group techniques, which indicates that the $R^2$-operator constitutes a relevant direction associated with an ultraviolet fixed point \cite{Lauscher:2002sq,Machado:2007ea,Codello:2008vh,Benedetti:2009rx,Benedetti:2012dx,Falls:2014tra,Falls:2018ylp,Falls:2020qhj}.\\
The present paper paves the way for a numerical study of the phase diagram of causal sets with an $R^2$ operator.

Our work also motivates another new direction, namely the construction of new operators that converge to curvature-invariants beyond the Ricci-scalar, e.g., $R_{\mu\nu} R^{\mu\nu}$ and $R_{\mu\nu\alpha\beta} R^{\mu\nu\alpha\beta}$. While the definition of objects with ``open indices'' from causal sets is a challenging task, because it requires the notion of a tangent space, curvature invariants are scalars and can therefore (in principle) be constructed from the causal set. In practise, Riemann invariants appear
in an expansion in Riemann normal coordinates, which is used as parts of the procedure of taking the $\ell\to 0$ limit of the discrete operators $B$. Thus, we expect that definitions of new discrete operators, based on  expressions inspired by Eq.~\eqref{eq:B-Operator_Def}, but with, for instance, different coefficients and different number of layers, can give us information about other curvature-invariants in a causal set.

\section*{Acknowledgments}
This work is supported by a research grant (29405) from VILLUM fonden. 

\appendix

\section{Expectation value of discrete operators \label{app:ExpctValue}}

In this appendix, we present some details on the derivation of the expectation value of discrete operators defined in causal sets.

In the construction of causal sets in terms of a sprinkling process, the distribution of points is given by the Poisson distribution, for which the probability to find $n$ points in a volume $V$ at sprinkling density $\rho$ is given by
\begin{eqnarray}
	P_n(V) = \frac{1}{n!}\left( \rho V\right)^n e^{-\rho\, V}.
\end{eqnarray}

We imagine spacetime to be discretized by small cells, labelled by the index $I$ and of spacetime volume $\delta V$, small compared to $1/\rho$, such that the probability of finding more than one element in one cell is essentially zero.
We then introduce two random variables, 
\begin{align}
	\chi_I = 
	\begin{cases}
		1 &, \quad \mbox{if $I$ contains an element},\\
		0 &, \quad \mbox{if $I$ contains no element}.
	\end{cases}
\end{align}
and
\begin{align}
	\tilde{\chi}_I^{(i)}(x) = 
	\begin{cases}
		1 &, \quad \mbox{if the causal volume between $x$ and $I$},\\
		& \hspace{.56cm} \mbox{contains $i-1$ elements},\\ 
		0 &, \quad \mbox{if the causal volume between $x$ and $I$ contains a}\\
		& \hspace{.56cm} \mbox{number of elements that is different from $i-1$}.
	\end{cases}
\end{align}
In this second definition, $I$ is understood to lie in the causal past of $x$.

The expectation value of $B(\phi(x))$ contains the expectation value of $\sum_{y\in L_i(x)} \phi(y)$. Using the definition of the random variables $\chi_I$ and $\tilde{\chi}_I^{(i)}(x)$, we can write
\begin{eqnarray}\label{eq:expt_value_Eq1}
	\left\langle\sum_{y \in L_i(x)}\phi(y) \right\rangle = 
	\bigg\langle\sum_I \chi_I\, \tilde{\chi}_I^{(i)}(x) \, \phi(y \in I) \bigg\rangle
\end{eqnarray}
This we can rewrite, using that the expectation value of a sum is the sum of expectation values and using that $\chi_I$ and $\tilde{\chi}_I^{(i)}(x)$ are independent, such that the expectation value of their product is the product of their expectation values, thus
\begin{eqnarray}\label{eq:expt_value_Eq2}
	\bigg\langle\sum_I \chi_I\, \tilde{\chi}_I^{(i)}(x) \, \phi(y \in I) \bigg\rangle 
	= \sum_I \langle \chi_I\rangle \langle \tilde{\chi}_I^{(i)}(x) \rangle \phi(y \in I).
\end{eqnarray}
Finally, we use the Poisson distribution to determine the expectation value of $\chi_I$ and $\tilde{\chi}_I^{(i)}(x)$. The probability for $\chi_I=1$ is given by $P(\chi_I=1)= \rho\, \delta V$ for small enough $\delta V$, where the exponential can be expanded to first order. At the same order in $\delta V$, $P(\chi_I=0)= 1-\rho\, \delta V$. Thus,
\begin{eqnarray}
	\langle \chi_I \rangle = 1\times P(\chi_I=1) + 0\times P(\chi_I=0) = \rho\,\delta V.
\end{eqnarray}
Further, we have that $P(\tilde{\chi}_I^{(i)}(x)=1) = {\frac{1}{(i-1)!}(\rho \,\mathcal{V}(x, y\in I))^{i-1} \,} e^{- \rho \, \mathcal{V}(x, y\in I)}$, which allows us to deduce that
\begin{equation}
	\begin{aligned}
		\langle \tilde{\chi}_I^{(i)}(x) \rangle &= 1\times P(\tilde{\chi}_I^{(i)}(x)=1) + 0 \times P(\tilde{\chi}_I^{(i)}(x)=0) \\
		&= \frac{(\rho \,\mathcal{V}(x, y\in I))^{i-1}}{(i-1)!} \, e^{-\rho\, \mathcal{V}(x,y\in I)}.
	\end{aligned}
\end{equation}
It follows that
	\begin{align}\label{eq:expt_value_Eq3}
		\sum_I \langle \chi_I\rangle \langle \tilde{\chi}_I^{(i)}(x) \rangle \phi(y \in I) =
		\rho \, \sum_I \delta V\, 
		\frac{(\rho \,\mathcal{V}(x, y\in I))^{i-1}}{(i-1)!} \, e^{-\rho\, \mathcal{V}(x,y\in I)}\,.
	\end{align}
	Combining Eqs.~\eqref{eq:expt_value_Eq1},~\eqref{eq:expt_value_Eq2} and \eqref{eq:expt_value_Eq3} and taking the limit $\delta V\to0$, we find 
	\begin{eqnarray}
		\left\langle\sum_{y \in L_i(x)}\phi(y) \right\rangle 
		= \rho \int_{J^-(x)} \hspace*{-.6cm}  d^dy\sqrt{-g(y)} \,  \frac{(\rho\, \mathcal{V}(x,y))^{i-1}}{(i-1)!}  \, e^{- \rho\mathcal{V}(x,y)} \phi(y) \, .
	\end{eqnarray}

We can proceed with similar arguments in the evaluation of the expectation value of $B^2(\phi(x))$. From Eq.~\eqref{eq:Bsq_extendend}, we can conclude that the expectation value of $B^2(\phi(x))$ contains $\left\langle\sum_{y\in L_i(x)} \sum_{z\in L_j(y)} \phi(z) \right\rangle$. Using the random variables $\chi$ and $\tilde{\chi}$, we can write
\begin{equation}\label{eq:exp_value_doublesum}
	\begin{aligned}
		&\hspace{-.4cm}\left\langle\sum_{y\in L_i(x)} \sum_{z\in L_j(y)} \phi(z) \right\rangle 
		= \left\langle\sum_{I} \sum_{I' \prec I} \chi_I \, \chi_{I'} \, \tilde{\chi}_I^{(i)}(x) \, \tilde{\chi}_{I'}^{(j)}(y\in I)  \, \phi(z\in I') \right\rangle  \\
		&\hspace{-.4cm}\quad = \sum_{I} \sum_{I'\prec I} P\left((\chi_I=1)\, {\rm and}\, (\chi_{I'}=1)\, {\rm and}\,( \tilde{\chi}_I^{(i)}=1)\, {\rm and}\,( \tilde{\chi}_{I'}^{(j)}=1) \right) \phi(z\in I').  
	\end{aligned} 
\end{equation}
Here, we have restricted $I'$ to lie in the causal past of $I$ in order for the rhs to equal the sum over layers on the lhs.

\begin{figure}[!t]
\begin{center}
\includegraphics[width=0.66\linewidth,clip=true, trim=22cm 15cm 22cm 0cm]{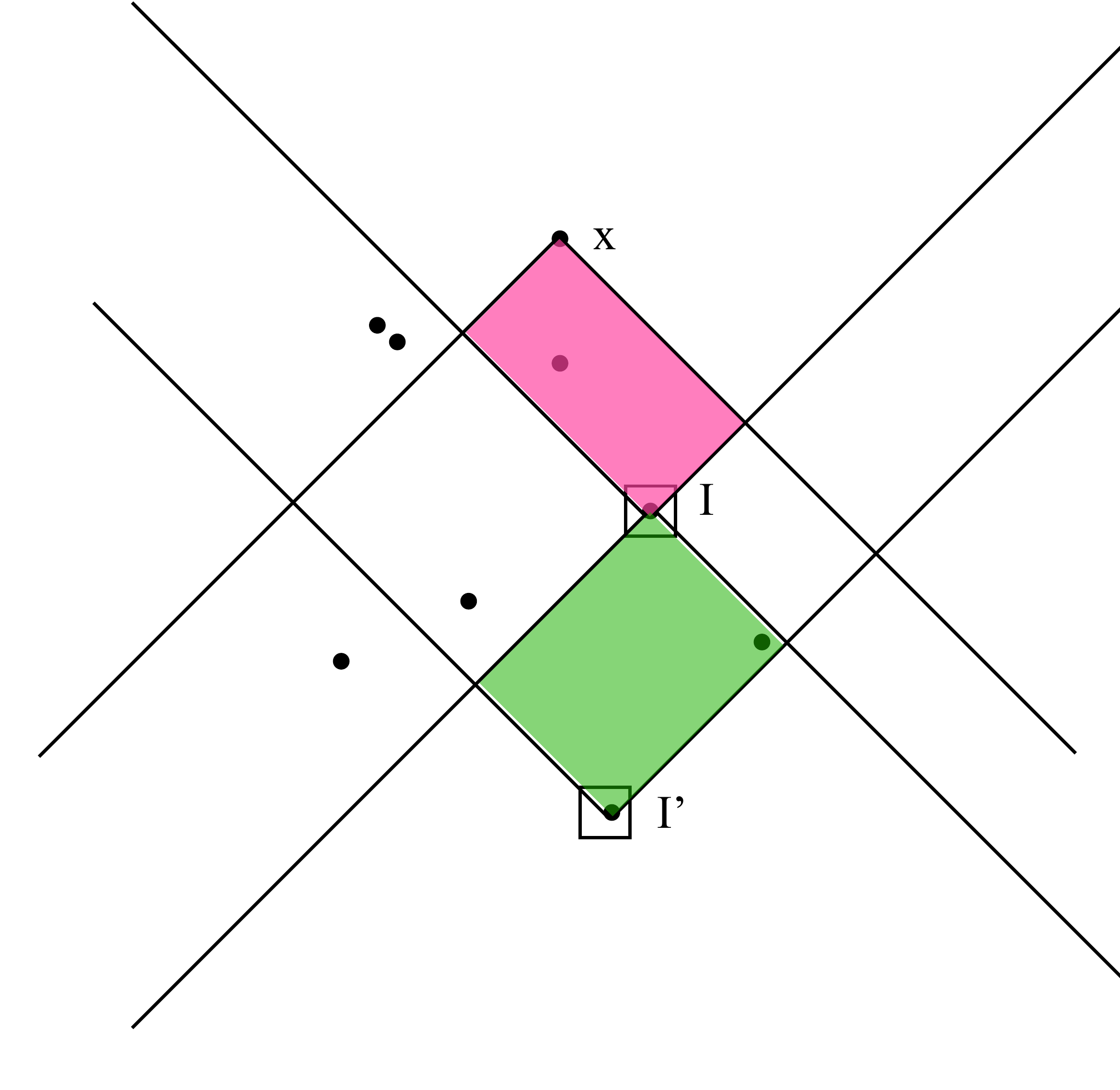}
\end{center}
\caption{\label{fig:illustrationdoublelayer} We show an example of one term in the sum in Eq.~\eqref{eq:exp_value_doublesum}. The probability to find a given number of points in each of the four volumes (namely, the infinitesimal volumes of the cells $I$ and $I'$ and the causal volume $\mathcal{V}(x,I)$ and $\mathcal{V}(I, I')$ is independent, because the volumes are non-overlapping in the limit of infinitesimal cell size of $I$ and $I$.\\
The figure also illustrates another aspect, namely that the causal volume between $x$ and $I'$ is not relevant for the construction. $B^2$ is therefore not a simple extension of $B$ by more causal layers into the past.}
\end{figure}

In the second step, we have already used that each of the random variables takes values 0 or 1 only and thus the expectation value is simply given to the probability that each random variable takes the value 1 simultaneously. Next, we factorize this probability into four independent probabilities. Here, we use that for a Poisson process, the probability to have $n$ elements in a volume $V$ depends only on $n$ and $V$ and is therefore independent of how many elements there are in another volume $V'$ which has no overlap with $V$. Accordingly, it holds that 
\begin{eqnarray}
	P_{n+n'}(V+V')\big|_{V \cap V' = \emptyset; n\, {\rm elements\, in}\, V} = P_n(V) \cdot P_{n'}(V').
\end{eqnarray}
In Eq.~\eqref{eq:exp_value_doublesum}, we have exactly such a situation, where we are considering four non-overlapping volumes and the number of elements within them; cf.~Fig.~\ref{fig:illustrationdoublelayer}.

Accordingly, we have that
\begin{eqnarray}
	\begin{aligned}
		&\sum_{I} \sum_{I'{\prec I}} P\left((\chi_I=1)\, {\rm and}\, (\chi_{I'}=1)\, {\rm and}\,( \tilde{\chi}_I^{(i)}=1)\, {\rm and}\,( \tilde{\chi}_{I'}^{(j)}=1) \right)\\
		&\qquad= P\left((\chi_I=1)\right) \cdot P\left((\chi_{I'}=1)\right)\cdot P\left((\tilde{\chi}_{I}^{(i)}=1)\right)\cdot P\left((\tilde{\chi}_{I'}^{(j)}=1)\right)\\
		&\qquad= \left(\rho \delta V \right)^2 \sum_I \sum_{I'\prec I} \left(  \frac{1}{(i-1)!}(\rho \mathcal{V}(x,y\in I))^{(i-1)}e^{-\rho \mathcal{V}(x,y\in I)} \right. \\
		&\quad\qquad\qquad\qquad\quad\quad\quad\times \left. \frac{1}{(j-1)!}(\rho \mathcal{V}(y\in I, z\in I'))^{j-1}e^{- \rho  \mathcal{V}(y\in I, z\in I')} \right) .
	\end{aligned}
\end{eqnarray}

Taking the limit $\delta V \to 0$, we find
\begin{equation}
	\begin{aligned}
		\left\langle\sum_{y\in L_i(x)} \sum_{z\in L_j(y)} \phi(z) \right\rangle 
		= \rho^2 &\int_{J^-(x)} \hspace*{-.6cm}  d^dy\sqrt{-g(y)} \,  \Bigg( \frac{(\rho\, \mathcal{V}(x,y))^{i-1}}{(i-1)!}  \, e^{- \rho\mathcal{V}(x,y)}  \\
		 \times &\int_{J^-(y)} \hspace*{-.6cm}  d^dz\sqrt{-g(z)} \,  \frac{(\rho\, \mathcal{V}(y,z))^{j-1}}{(j-1)!}  \, e^{- \rho\mathcal{V}(y,z)} \, \phi(z) \Bigg) \, .
	\end{aligned}
\end{equation}
	
We can generalize the last result in the following way
\begin{equation}
	\begin{aligned}
		&\left\langle\sum_{y_1\in L_{i_1}(x)} \cdots \hspace*{-.3cm}
		\sum_{y_n\in L_{i_n}(y_{n-1})} \hspace*{-.4cm} \phi(y_n) \right\rangle =  \\
		&\qquad = \rho^n\, 
		\int_{J^-(x)} \hspace*{-.6cm}  d^dy_1\sqrt{-g(y_1)} \,\cdots\,
		\int_{J^-(y_{n-1})} \hspace*{-.6cm}  d^dy_n\sqrt{-g(y_n)}  \\
		&\qquad 
		\times \frac{(\rho\, \mathcal{V}(x,y_1))^{i_1-1}}{(i_1-1)!} \,e^{- \rho\mathcal{V}(x,y_1)} 
		\,\cdots\,
		\frac{(\rho\, \mathcal{V}(y_{n-1},y_n))^{i_n-1}}{(i_n-1)!} \,e^{- \rho\mathcal{V}(y_{n-1},y_n)}  \,\, \phi(y_n) \,,
	\end{aligned}
\end{equation}
which implies the recursive relation
\begin{align}
	&\left\langle\sum_{y_1\in L_{i_1}(x)} \cdots \hspace*{-.3cm}
	\sum_{y_n\in L_{i_n}(y_{n-1})} \hspace*{-.4cm} \phi(y_n) \right\rangle =  \\
	&\quad=
	\rho\, 
	\int_{J^-(x)} \hspace*{-.6cm}  d^dy_1\sqrt{-g(y_1)}
	\frac{(\rho\, \mathcal{V}(x,y_1))^{i_1-1}}{(i_1-1)!} \,e^{- \rho\mathcal{V}(x,y_1)} 
	\left\langle\sum_{y_2\in L_{i_2}(y_1)} \cdots \hspace*{-.3cm}
	\sum_{y_n\in L_{i_n}(y_{n-1})} \hspace*{-.4cm} \phi(y_n) \right\rangle \,. \nn
\end{align}
We have used this recursive relation in Sec. \ref{sec:higher_order_operators} to derive derive the continuum limit of the operator $B^n(\phi(x))$.

\bibliographystyle{unsrt}
\bibliography{refs.bib}

\end{document}